\documentstyle[12pt,procsla]{article}
\begin{document}


\begin{titlepage} 

\vspace*{1.5truecm}

\begin{flushright}
CERN-TH/97-295\\
hep-ph/9710455
\end{flushright}
 
\vspace{2cm}
 
\begin{center}
\Large\bf CP Violation and Strategies for\\
\vspace{0.3truecm}
 Extracting CKM Phases
\end{center}

\vspace{1.2cm}
 
\begin{center}
Robert Fleischer\\
{\sl Theory Division, CERN, CH-1211 Geneva 23, Switzerland}
\end{center}
 
\vspace{1.5cm}

\begin{center}
{\bf Abstract}\\[0.3cm]
\parbox{12.5cm}{
The phenomenon of CP violation and strategies for extracting the 
angles of the unitarity triangle are reviewed. Special emphasis is given 
to the $B$-meson system, as it plays an outstanding role to test the 
Standard Model description of CP violation at future experimental $B$-physics 
facilities. Both general aspects and recent developments are discussed. 
It is pointed out that combined branching ratios for $B_{u,d}\to\pi K$ modes, 
which have been observed recently by the CLEO collaboration, and untagged
$B_s\to K\overline{K}$ decays may allow us to derive interesting constraints 
on the CKM angle $\gamma$ and may eventually lead to a solid determination
of this angle.}
\end{center}
 
\vspace{2cm}
 
\begin{center}
{\sl Invited plenary talk given at the\\
IVth International Workshop on Progress in Heavy Quark Physics\\
Rostock, Germany, 20--22 September 1997\\
To appear in the Proceedings}
\end{center}
 
\vspace{1.8cm}
 
\vfil
\noindent
CERN-TH/97-295\\
October 1997
 
\end{titlepage}
 
\thispagestyle{empty}
\vbox{}
\newpage
 
\setcounter{page}{1}
 

\begin{center}
{\bf CP VIOLATION AND STRATEGIES FOR EXTRACTING CKM PHASES}

\vspace*{1cm}
ROBERT FLEISCHER\\
{\it Theory Division, CERN\\ 
CH-1211 Geneva 23, Switzerland\\}
\end{center}

\vspace*{0.85cm}
\begin{abstracts}
{\small The phenomenon of CP violation and strategies for extracting the 
angles of the unitarity triangle are reviewed. Special emphasis is given 
to the $B$-meson system, as it plays an outstanding role to test the 
Standard Model description of CP violation at future experimental $B$-physics 
facilities. Both general aspects and recent developments are discussed. 
It is pointed out that combined branching ratios for $B_{u,d}\to\pi K$ modes, 
which have been observed recently by the CLEO collaboration, and untagged
$B_s\to K\overline{K}$ decays may allow us to derive interesting constraints 
on the CKM angle $\gamma$ and may eventually lead to a solid determination
of this angle.}
\end{abstracts}

\vspace*{0.25cm}
\section{Introduction}\label{intro}
CP violation plays a central and fundamental role in present particle 
physics. Since it is one of the least experimentally explored 
phenomena of the Standard Model, it may still shed light on new physics.
To this end, it is crucial to search for CP-violating processes that 
can be analysed in a clean way within the Standard Model framework.

Within the Standard Model of electroweak interactions, CP violation
is closely related to the Cabibbo--Kobayashi--Maskawa matrix\,\cite{ckm}
(CKM matrix) connecting the electroweak eigenstates of the $d$, $s$ and
$b$ quarks with their mass eigenstates. As far as CP violation is concerned,
the central feature of the CKM matrix is that -- in addition to three 
generalized Cabibbo-type angles -- also a {\it complex phase} is needed in 
the three-generation case to parametrize the CKM matrix. This complex phase 
is the origin of CP violation within the Standard Model.
  
A closer look shows that CP-violating observables are proportional to the 
following combination of CKM matrix elements: 
\begin{equation}
J_{\rm CP}=\pm\,\mbox{Im}\left(V_{i\alpha}V_{j\beta}V_{i\beta}^\ast 
V_{j\alpha}^\ast\right)\quad(i\not=j,\,\alpha\not=\beta)\,,
\end{equation}
which represents a measure of the ``strength'' of CP violation within the 
Standard Model\,\cite{jarlskog}. Since $J_{\rm CP}={\cal O}(10^{-5})$, CP 
violation is a small effect. Typically several new complex couplings are 
present in many scenarios of new physics\,\cite{new-phys}, yielding additional 
sources for CP violation.

Concerning phenomenological applications, the parametrization 
\begin{equation}\label{wolf2}
\hat V_{\mbox{{\scriptsize CKM}}} =\left(\begin{array}{ccc}
1-\frac{1}{2}\lambda^2 & \lambda & A\lambda^3 R_b\, e^{-i\gamma} \\
-\lambda & 1-\frac{1}{2}\lambda^2 & A\lambda^2\\
A\lambda^3R_t\,e^{-i\beta} & -A\lambda^2 & 1
\end{array}\right)+\,{\cal O}(\lambda^4)
\end{equation}
with $\lambda=0.22$ and
\begin{equation}\label{RbRt}
A\equiv\frac{1}{\lambda^2}\left|V_{cb}\right|=0.81\pm0.06
\end{equation}
\begin{equation}
R_b\equiv\frac{1}{\lambda}\left|\frac{V_{ub}}{V_{cb}}\right|=
\sqrt{\rho^2+\eta^2}=0.36\pm0.08,\quad
R_t\equiv\frac{1}{\lambda}\left|\frac{V_{td}}{V_{cb}}\right|=
\sqrt{(1-\rho)^2+\eta^2}={\cal O}(1)
\end{equation}
turns out to be very useful. This parametrization is a modification of the 
Wolfenstein parametrization\,\cite{wolf} expliciting not only the hierarchy 
of the CKM elements, but also the dependence on the angles 
$\beta=\beta(\rho,\eta)$ and $\gamma=\gamma(\rho,\eta)$ of the 
usual ``non-squashed'' unitarity triangle of the CKM matrix\,\cite{ut}.

\section{A Brief Look at CP Violation in Kaon Decays}\label{kaon-sys}
So far, CP violation has been observed only within the neutral $K$-meson 
system, although the discovery of this phenomenon goes back to the year 
1964\,\cite{ccft}. In the neutral kaon system, CP violation is described by 
two complex quantities, called $\varepsilon$ and $\varepsilon'$, which are 
defined by the following ratios of decay amplitudes:
\begin{equation}\label{defs-eps}
\frac{A(K_{\rm L}\to\pi^+\pi^-)}{A(K_{\rm S}
\to\pi^+\pi^-)}=\varepsilon+\varepsilon',\quad
\frac{A(K_{\rm L}\to\pi^0\pi^0)}{A(K_{\rm S}
\to\pi^0\pi^0)}=\varepsilon-2\varepsilon'.
\end{equation}
While $\varepsilon=(2.280\pm0.013)\times e^{i\frac{\pi}{4}}\times 10^{-3}$
parametrizes ``indirect'' CP violation originating from the fact that
the mass eigenstates of the neutral $K$-meson system are not CP eigenstates,
the quantity Re$(\varepsilon'/\varepsilon)$ measures ``direct'' CP violation 
in $K\to\pi\pi$ transitions. The CP-violating observable $\varepsilon$ plays 
an important role to constrain the unitarity triangle\,\cite{bf-rev} and 
informs us in particular about a positive value of the
Wolfenstein parameter~$\eta$.

Despite enormous efforts, the experimental situation concerning 
Re$(\varepsilon'/\varepsilon)$ is still unclear at present. Whereas the 
CERN experiment NA31 finds Re$(\varepsilon'/\varepsilon)=
(23\pm7)\times 10^{-4}$, which already indicates direct CP violation, 
the result Re$(\varepsilon'/\varepsilon)=(7.4\pm5.9)\times 10^{-4}$ of the 
Fermilab experiment E731 provides no unambiguous evidence for a non-zero 
effect. These results have been published already in 1993. In the near 
future, this unsatisfactory experimental situation will hopefully be clarified
by improved measurements at CERN and Fermilab, as well as by the KLOE 
experiment at DA$\Phi$NE. 

From a theoretical point of view, Re$(\varepsilon'/\varepsilon)$ is 
unfortunately not in much better shape. The corresponding calculations are 
very involved and suffer at present from large hadronic 
uncertainties\,\cite{bf-rev}. 
Consequently, that observable will not allow a powerful test of the 
CP-violating sector of the Standard Model, unless the hadronic matrix elements 
of the relevant operators can be brought under better control. Probably the 
major goal of a possible future observation of Re$(\varepsilon'/\varepsilon)
\not=0$ would hence be the unambiguous exclusion of ``superweak'' models of 
CP violation\,\cite{superweak}.
 
In respect of testing the Standard Model description of CP violation, the 
rare decays $K_{\rm L}\to\pi^0\nu\overline{\nu}$ and $K^+\to\pi^+
\nu\overline{\nu}$ are more promising. These decays -- in particular the first 
one -- are very clean from a theoretical point of view\,\cite{bb-nlo}. 
The branching ratios of these decays allow a determination 
of the unitarity triangle, if the top quark mass $m_t$ and the CKM element 
$|V_{cb}|$ are used as an additional input. A detailed analysis shows that 
in particular $\sin(2\beta)$ can be extracted with respectable 
accuracy\,\cite{bb}. Thus, one has a powerful tool to probe physics beyond the 
Standard Model by comparing the value of $\sin(2\beta)$ determined in
this way with the one extracted from CP violation in the ``gold-plated'' 
mode $B_d\to J/\psi\, K_{\rm S}$ (see Subsection~3.1). Although the first 
experimental evidence for $K^+\to\pi^+\nu\overline{\nu}$ has recently
been reported by the E787 collaboration\,\cite{E787}, measurements
of these rare kaon decays -- in particular of $K_{\rm L}\to\pi^0
\nu\overline{\nu}$ -- are very challenging. Within the Standard Model, the
expected branching ratios\,\cite{bf-rev,bb-nlo} are 
BR$(K^+\to\pi^+\nu\overline{\nu})=(9.1\pm3.6)\times10^{-11}$ and 
BR$(K_{\rm L}\to\pi^0\nu\overline{\nu})=(2.8\pm1.7)\times10^{-11}$.
Nevertheless, there are plans to explore these important decays at BNL, 
FNAL and KEK.

To summarize the status of the kaon system, the observed CP violation in 
$K$-meson decays can be described successfully by the Standard Model at 
present. This feature is, however, not surprising, since so far only a single 
CP-violating observable, $\varepsilon$, has to be fitted. Consequently
many different non-Standard Model descriptions of CP violation are 
imaginable\,\cite{new-phys}. From the brief discussion given above, it is 
obvious that the kaon system by itself cannot provide the whole picture of 
CP violation. Therefore it is essential to study CP violation outside this 
system. In this respect, the $B$ system appears to be most promising, which 
is also reflected by the tremendous experimental efforts at future 
$B$-factory facilities. There are of course also other interesting systems to 
investigate CP violation and to search for signals of new physics, for
instance the $D$-meson system\,\cite{D-rev}, where sizeable mixing or 
CP-violating effects would signal new physics because of the tiny Standard 
Model ``background''. In this presentation, I cannot discuss these systems 
in more detail and shall focus on $B$ decays.

\section{CP Violation in B Decays}\label{CP-B}
The $B$-meson system is expected to provide a very fertile ground for
testing the Standard Model description of CP violation. In this respect,
the major role is played by non-leptonic $B$ decays, which can be divided 
into three decay classes: decays receiving both tree and penguin 
contributions, pure tree decays, and pure penguin decays. A distinction is
made between two types of penguin topologies: gluonic (QCD) and electroweak 
(EW) penguins related to strong and electroweak interactions, respectively. 
Because of the large top quark mass, also the latter operators play an 
important role in several processes\,\cite{rev}.

In order to analyse non-leptonic $B$ decays theoretically, one uses 
low-energy effective Hamiltonians, which are calculated by making use of the 
operator product expansion, yielding transition matrix elements of the
following structure:
\begin{equation}\label{ee2}
\langle f|{\cal H}_{\rm eff}|i\rangle\propto\sum_k C_k(\mu)
\langle f|Q_k(\mu)|i\rangle\,.
\end{equation}
The operator product expansion allows us to separate the short-distance
contributions to Eq.\ (\ref{ee2}) from the long-distance contributions, which
are described by perturbative Wilson coefficient functions $C_k(\mu)$
and non-perturbative hadronic matrix elements $\langle f|Q_k(\mu)|
i\rangle$, respectively. As usual, $\mu$ denotes an appropriate
renormalization scale. 

In the case of $|\Delta B|=1$, $\Delta C=\Delta U=0$ transitions, which play
the major role in the following discussion, we have
\begin{equation}\label{e3}
{\cal H}_{\mbox{{\scriptsize eff}}}={\cal H}_{\mbox{{\scriptsize 
eff}}}(\Delta B=-1)+{\cal H}_{\mbox{{\scriptsize eff}}}(\Delta B=-1)^\dagger
\end{equation}
with
\begin{equation}\label{e4}
{\cal H}_{\mbox{{\scriptsize eff}}}(\Delta B=-1)=\frac{G_{\mbox{{\scriptsize 
F}}}}{\sqrt{2}}\left[\sum\limits_{j=u,c}V_{jq}^\ast V_{jb}\left\{\sum
\limits_{k=1}^2Q_k^{jq}\,C_k(\mu)+\sum\limits_{k=3}^{10}Q_k^{q}\,C_k(\mu)
\right\}\right].
\end{equation}
Here $\mu={\cal O}(m_b)$, $Q_k^{jq}$ are four-quark operators, and the label
$q\in\{d,s\}$ corresponds to $b\to d$ and $b\to s$ transitions. 
The index $k$ distinguishes between ``current--current'' 
$(k\in\{1,2\})$, QCD $(k\in\{3,\ldots,6\})$ and electroweak 
$(k\in\{7,\ldots,10\})$ penguin operators, which are related to tree-level, 
QCD and electroweak penguin processes, respectively. The evaluation 
of such low-energy effective Hamiltonians has been reviewed in 
Ref.\,\cite{bbl-rev}, where the four-quark operators are given explicitly 
and numerical values for their Wilson coefficient functions can be found.

\subsection{CP Asymmetries in Neutral B Decays}\label{CP-asym-neut}
A particularly simple and interesting situation arises, if we restrict 
ourselves to decays of neutral $B_q$ mesons ($q\in\{d,s\}$) into CP 
self-conjugate final states $|f\rangle$ satisfying the relation 
$({\cal CP})|f\rangle=\pm|f\rangle$. In that case the corresponding 
time-dependent CP asymmetry can be expressed as
\begin{eqnarray}
\lefteqn{a_{\mbox{{\scriptsize CP}}}(t)\equiv\frac{\Gamma(B^0_q(t)\to f)-
\Gamma(\overline{B^0_q}(t)\to f)}{\Gamma(B^0_q(t)\to f)+
\Gamma(\overline{B^0_q}(t)\to f)}=}\nonumber\\
&&{\cal A}^{\mbox{{\scriptsize dir}}}_{\mbox{{\scriptsize CP}}}(B_q\to f)
\cos(\Delta M_q\,t)+{\cal A}^{\mbox{{\scriptsize
mix--ind}}}_{\mbox{{\scriptsize CP}}}(B_q\to f)\sin(\Delta M_q\,t)
\,,\label{ee6}
\end{eqnarray}
where the direct CP-violating contributions have been separated from
the mixing-induced CP-violating contributions, which are characterized by
\begin{equation}\label{ee7}
{\cal A}^{\mbox{{\scriptsize dir}}}_{\mbox{{\scriptsize CP}}}(B_q\to f)\equiv
\frac{1-\left|\xi_f^{(q)}\right|^2}{1+\left|\xi_f^{(q)}\right|^2}\quad
\mbox{and}\quad
{\cal A}^{\mbox{{\scriptsize mix--ind}}}_{\mbox{{\scriptsize
CP}}}(B_q\to f)\equiv\frac{2\,\mbox{Im}\,\xi^{(q)}_f}{1+\left|\xi^{(q)}_f
\right|^2}\,,
\end{equation}
respectively. Here direct CP violation refers to CP-violating effects
arising directly in the corresponding decay amplitudes, whereas mixing-induced
CP violation is related to interference between 
$B_q^0$--$\overline{B_q^0}$ mixing and decay processes. Note that the 
expression Eq.~(\ref{ee6}) has to be modified in the $B_s$ case 
for $t\mathrel{\hbox{\rlap{\hbox{\lower4pt\hbox{$\sim$}}}\hbox{$>$}}}1/
|\Delta\Gamma_s|$ because of the expected sizeable width difference 
$\Delta\Gamma_s$\,\cite{dun}.

In general, the observable
\begin{equation}
\xi_f^{(q)}\equiv e^{-i\phi_{\mbox{{\scriptsize M}}}^{(q)}}
\frac{A(\overline{B^0_q}\to f)}{A(B_q\to f)}\,,\quad\mbox{where}\quad
\phi_{\mbox{{\scriptsize M}}}^{(q)}=\left\{\begin{array}{cr}
2\beta&\mbox{for $q=d$}\\
0&\mbox{for $q=s$}\end{array}\right. 
\end{equation}
denotes the weak $B_q^0$--$\overline{B_q^0}$ mixing 
phase, suffers from large hardonic uncertainties, which are introduced through 
the decay amplitudes $A$. There is, however, a very important special case, 
where these uncertainties cancel. It is given if $B_q\to f$ is dominated by 
a single CKM amplitude. In that case, $\xi_f^{(q)}$ takes the simple form
\begin{equation}\label{ee10}
\xi_f^{(q)}=\mp\exp\left[-i\left(\phi_{\mbox{{\scriptsize M}}}^{(q)}-
\phi_{\mbox{{\scriptsize D}}}^{(f)}\right)
\right],
\end{equation}
where $\phi_{\mbox{{\scriptsize D}}}^{(f)}$ is a characteristic weak decay
phase, which is given by
\begin{equation}\label{e11}
\phi_{\mbox{{\scriptsize D}}}^{(f)}=\left\{\begin{array}{cc}
-2\gamma&\mbox{for dominant $\bar b\to\bar u\,u\,\bar r$ CKM amplitudes
in $B_q\to f$}\\
0&\,\mbox{for dominant $\bar b\to\bar c\,c\,\bar r\,$ CKM amplitudes
in $B_q\to f$.}
\end{array}\right.
\end{equation}
The label $r\in\{d,s\}$ distinguishes between $\bar b\to\bar d$ and 
$\bar b\to\bar s$ transitions.

Important applications and well-known examples of this formalism are the 
decays $B_d\to J/\psi\, K_{\mbox{{\scriptsize S}}}$ and $B_d\to\pi^+\pi^-$.
If one goes through the relevant Feynman diagrams contributing to the former
channel (for a detailed recent discussion, see Ref.\,\cite{rev}), one finds 
that it is dominated to excellent accuracy by the $\bar b\to\bar cc\bar s$ 
CKM amplitude. Therefore the decay phase vanishes and we have
\begin{equation}\label{e12}
{\cal A}^{\mbox{{\scriptsize mix--ind}}}_{\mbox{{\scriptsize
CP}}}(B_d\to J/\psi\, K_{\mbox{{\scriptsize S}}})=+\sin[-(2\beta-0)]\,.
\end{equation}
Since Eq.~(\ref{ee10}) applies to excellent accuracy to the decay
$B_d\to J/\psi\, K_{\mbox{{\scriptsize S}}}$ -- the point is that penguins
enter essentially with the same weak phase as the leading tree
contribution -- it is usually referred to as the ``gold-plated'' mode
to measure the CKM angle $\beta$\,\cite{csbs}.

In the case of $B_d\to\pi^+\pi^-$, mixing-induced CP violation would 
measure $-\sin(2\alpha)$ in a clean way through
\begin{equation}\label{e13}
{\cal A}^{\mbox{{\scriptsize mix--ind}}}_{\mbox{{\scriptsize
CP}}}(B_d\to\pi^+\pi^-)=-\sin[-(2\beta+2\gamma)]=-\sin(2\alpha)\,,
\end{equation}
if there were no penguin contributions present. However, such contributions
are there and destroy the cleanliness of Eq.~(\ref{e13}). The 
corresponding hadronic uncertainties were discussed by many authors in the 
literature. 
There are even methods to control these uncertainties in a quantitative 
way. Unfortunately, these strategies are usually rather challenging in 
practice. The most important examples are the $B\to\pi\pi$ isospin 
triangles proposed by Gronau and London\,\cite{gl}, and an approach 
using $B\to\rho\,\pi$ modes\,\cite{Brhopi}. An approximate method to 
correct for the penguin uncertainties in $B_d\to\pi^+\pi^-$ that appears 
to be promising for the early days of the $B$-factory era was proposed 
in Ref.\,\cite{fm1}. For a detailed discussion of these and other strategies, 
the reader is referred to reviews on this topic, for instance
Refs.\,\cite{rev,cp-revs}. 
 
A decay appearing frequently as a tool to determine the CKM angle $\gamma$ 
is $B_s\to\rho^0 K_{\rm S}$. In that case, however, 
penguins are expected to lead to serious problems -- more serious than in
$B_d\to\pi^+\pi^-$ -- so that this mode appears to be the ``wrong'' way to 
extract $\gamma$\,\cite{rev}. Other strategies allowing meaningful 
determinations of this angle will be discussed below. 

\subsection{CP Violation in Non-leptonic Penguin Modes as a Probe of
New Physics}
In order to test the Standard Model description of CP violation, 
penguin-induced modes play an important role. Because of the loop-suppression
of these ``rare'' processes, it is possible -- and indeed it is the case in
several specific model calculations -- that new-physics contributions to 
these decays are of similar magnitude as those of the 
Standard Model\,\cite{new-phys}. An example is the $b\to d$ penguin mode 
$B_d\to K^0\overline{K^0}$ (for an analysis of new-physics effects, see 
Ref.\,\cite{mpw}). If one assumes that penguins with internal top quarks 
play the dominant role in this decay, the weak $B^0_d$--$\overline{B^0_d}$ 
mixing and $B_d\to K^0\overline{K^0}$ decay phases cancel in the 
corresponding observable $\xi^{(d)}_{K^0\overline{K^0}}$, implying 
{\it vanishing} CP violation in that decay. Consequently one would conclude 
that a measurement of non-vanishing CP violation in $B_d\to K^0\overline{K^0}$
would signal physics beyond the Standard Model. However, long-distance 
effects related to penguins with internal charm and up quarks may easily
spoil the assumption of top quark dominance\,\cite{rev,bf1}. These 
contributions may lead to sizeable CP violation in $B_d\to K^0\overline{K^0}$ 
even within the Standard Model\,\cite{my-KK.bar}, so that a measurement 
of such CP asymmetries would not necessarily imply new physics, as claimed in 
several previous papers. Unfortunately a measurement of these effects will 
be difficult, since the Standard Model expectation for the corresponding 
branching ratio is ${\cal O}(10^{-6})$, which is still one order of magnitude 
below the recent CLEO bound\,\cite{cleo} 
$\mbox{BR}(B_d\to K^0\overline{K^0})<1.7\times10^{-5}$. 

More promising in this respect and -- more importantly -- to search for 
physics beyond the Standard Model, is the $b\to s$ penguin mode $B_d\to\phi\, 
K_{\rm S}$. The branching ratio for this decay is expected to be of 
${\cal O}(10^{-5})$ and may be large enough to investigate this channel 
at future $B$ factories. Interestingly there is, to a very good 
approximation, no non-trivial CKM phase present in the corresponding decay 
amplitude\,\cite{rev}, so that direct CP violation vanishes and mixing-induced 
CP violation measures simply the weak $B^0_d$--$\overline{B^0_d}$ mixing 
phase. This statement does {\it not} require the questionable 
assumption of top quark dominance in penguin amplitudes. Consequently, an 
important probe for new physics in $b\to s$ flavour-changing neutral-current
processes is provided by the relation
\begin{equation}
{\cal A}^{\mbox{{\scriptsize mix--ind}}}_{\mbox{{\scriptsize
CP}}}(B_d\to J/\psi\, K_{\mbox{{\scriptsize S}}})={\cal 
A}^{\mbox{{\scriptsize mix--ind}}}_{\mbox{{\scriptsize
CP}}}(B_d\to \phi\, K_{\mbox{{\scriptsize S}}})=-\sin(2\beta)\,,
\end{equation}
which holds within the Standard Model framework. The theoretical accuracy 
of this relation is limited by certain neglected terms that are 
CKM-suppressed by ${\cal O}(\lambda^2)$ and may lead to tiny direct 
CP-violating asymmetries in $B_d\to\phi\, K_{\rm S}$ of at most 
${\cal O}(1\%)$\,\cite{rev}. The importance of $B_d\to\phi\, K_{\rm S}$ and 
similar modes, such as $B_d\to\eta' K_{\rm S}$, to search for new physics 
in $b\to s$ transitions has recently been emphasized by several 
authors\,\cite{rev,BdPhiKs}.

\subsection{The $B_s$ System in the Light of $\Delta\Gamma_s$}\label{Bssys}
In the $B_s$ system very rapid $B^0_s$--$\overline{B^0_s}$ oscillations are
expected, requiring an excellent vertex resolution system. Studies of CP 
violation in $B_s$ decays are therefore regarded as being very difficult. An 
alternative route to investigate CP-violating effects may be provided by 
the width difference $\Delta\Gamma_s/\Gamma_s={\cal O}(20\%)$ arising from
CKM-favoured $b\to c\bar c s$ transitions into final states that are
common to both $B^0_s$ and $\overline{B^0_s}$. Because of this width 
difference, already {\it untagged} data samples of $B_s$ decays may exhibit 
CP-violating effects\,\cite{dun}. 
 
In the recent literature\,\cite{fd1,fd2} (for a review, see for instance 
Ref.\,\cite{hawaii}), several ``untagged strategies'' to extract the CKM 
angle $\gamma$ have been proposed, using for example angular distributions in 
$B_s\to K^{\ast+}K^{\ast-},\,K^{\ast0}\overline{K^{\ast0}}$ or 
$B_s\to D^{\ast}\phi,\,D_s^{\ast\pm}K^{\ast\mp}$ decays. To illustrate these 
strategies in more detail, let us focus on an approach to determine $\gamma$ 
from untagged $B_s\to K^0\overline{K^0}$ and $B_s\to K^+K^-$ decays that 
was proposed in Ref.\,\cite{fd1}. Neglecting tiny EW penguin contributions
and using the $SU(2)$ isospin symmetry of strong interactions, the 
corresponding untagged rates, which are defined by 
\begin{equation}\label{untagged}
\Gamma[f(t)]\equiv\Gamma(B_s^0(t)\to f)+\Gamma(\overline{B^0_s}(t)\to f)\,,
\end{equation}
take the form \cite{fd1,BpiKBsKK}
\begin{equation}\label{EE5}
\Gamma[K^0\overline{K^0}(t)]={\cal C}\, |P'_s|^2\,e^{-\Gamma_L^{(s)} t}\,,
\quad 
\Gamma[K^+K^-(t)]={\cal C}\, |P'_s|^2\left[a\,e^{-\Gamma_L^{(s)} t}+
b\,e^{-\Gamma_H^{(s)} t}\right],
\end{equation}
where ${\cal C}$ is a trivially calculable phase-space factor and
\begin{equation}\label{Def-ab}
a=1-2\,r_s\cos\delta_s\,\cos\gamma+r_s^2\cos^2\gamma\,,\quad
b=r_s^2\sin^2\gamma\,,
\end{equation}
with $r_s\equiv|T'_s|/|P'_s|$, $\delta_s\equiv\delta_{T'_s}-
\delta_{P'_s}$. The amplitudes $P'_s$ and $T'_s$ denote QCD penguin and 
colour-allowed $\bar b\to\bar uu\bar s$ current--current operator 
contributions, respectively. Since 
the untagged $B_s\to K^0\overline{K^0}$ rate measures ${\cal C}\,|P'_s|^2$, 
which gives the normalization of the untagged $B_s\to K^+K^-$ rate, both 
$a$ and $b$ can be determined. If the amplitude ratio $r_s$ is known, the 
CKM angle $\gamma$ and the strong phase $\delta_s$ can be extracted from 
these observables. Since $|P'_s|$ is measured through the untagged $B_s\to
K^0\overline{K^0}$ rate, this required input acutally corresponds to $|T'_s|$,
which can be fixed, e.g.\ through $B^+\to\pi^+\pi^0$ and $SU(3)$ flavour
symmetry, ``factorization'', or 
hopefully lattice gauge theory one day. The required information about $|T'_s|$
introduces some model-dependence into the extracted value of $\gamma$. 
However, as we will see in more detail in Subsection 3.6, it is possible to 
derive interesting {\it bounds} on $\gamma$ from the untagged 
$B_s\to K\overline{K}$ rates that do not suffer from such a 
model dependence\,\cite{BpiKBsKK}. 

Let me note that the $B_s$ system provides also interesting probes for 
physics beyond the Standard Model. Important examples are the decays 
$B_s\to D_s^+D_s^-$ and $B_s\to J/\psi\phi$. The latter is the counterpart of 
the ``gold-plated'' mode $B_d\to J/\psi K_{\rm S}$ to measure $\beta$. 
These transitions are dominated by a single CKM amplitude and allow -- 
in principle even from their untagged data samples\,\cite{fd1} -- the 
extraction of a CP-violating weak phase 
$\phi_{\rm CKM}\equiv2\lambda^2\eta$, which is expected to be of 
${\cal O}(0.03)$ within the Standard Model\,\cite{bf-rev}. Consequently, 
an extracted value of $\phi_{\rm CKM}$ that is much larger than this 
Standard Model expectation would signal new-physics contributions to
$B^0_s$--$\overline{B^0_s}$ mixing. The untagged
$B_s\to K^0\overline{K^0}$ rate is also interesting to search for new physics.
If future experiments should find that its time evolution depends -- in
contrast to Eq.~(\ref{EE5}) -- on {\it two} exponentials, we would have an 
indication for physics beyond the Standard Model\,\cite{BpiKBsKK}.

\subsection{Extracting CKM Angles with Amplitude Relations}
Since mixing effects are absent in the charged $B$ system, the measurement of
a non-vanishing CP asymmetry in a charged $B$ decay would give us unambiguous
evidence for direct CP violation, thereby ruling out ``superweak'' models.
Such CP asymmetries arise from interference between decay amplitudes with
both different CP-violating weak and CP-conserving strong phases. Whereas
the weak phases are related to the CKM matrix, the strong phases are 
induced by strong final-state interaction effects and introduce severe
theoretical uncertainties into the calculation, destroying in general the
clean relation of the CP asymmetry to the phases of the CKM matrix. 
 
Nevertheless, there are decays of charged $B$ mesons that play an important
role to extract angles of the unitarity triangle, in particular for $\gamma$.
In this case amplitude relations -- either exact or approximate ones 
based on flavour symmetries -- are used. A recent review of these methods can 
be found in Ref.\,\cite{rev}. The ``prototype'' is the approach to determine 
$\gamma$ with the help of triangle relations among the $B^\pm\to D K^\pm$ 
decay amplitudes proposed by Gronau and Wyler\,\cite{gw}. Unfortunately the 
corresponding triangles are expected to be very ``squashed'' ones. Moreover
one has to deal with additional experimental problems\,\cite{ads}, so that 
this approach is very difficult from a practical point of view. Recently 
more refined variants have been proposed by Atwood, Dunietz and 
Soni\,\cite{ads}.
 
About three years ago, several methods to extract CKM angles were presented
by Gronau, Hern\'andez, London and Rosner, who have combined the $SU(3)$
flavour symmetry of strong interactions with certain plausible dynamical 
assumptions to derive relations among $B\to\pi\pi,\pi K,K\overline{K}$ decay 
amplitudes\,\cite{ghlr}. This approach has been very popular over the recent 
years and requires only a measurement of the relevant branching 
ratios. A closer look shows, however, that it has -- despite its 
attractiveness -- theoretical limitations:
the $SU(3)$ relations are not valid exactly, QCD penguins 
with internal charm and up quarks play an important role in several cases, 
and interestingly also EW penguins lead to complications. In order to eliminate
the EW penguin contributions, usually very involved strategies are needed. 
A detailed discussion of all these methods (mainly to extract $\gamma$) is 
beyond the scope of this presentation and the reader is referred to a recent 
review\,\cite{rev} and references therein. 

\subsection{Constraining $\gamma$ with $B_{u,d}\to\pi K$ Modes}
A simple approach to determine $\gamma$ with the help of the branching ratios
for $B^+\to\pi^+K^0$, $B^0_d\to\pi^-K^+$ and their charge-conjugates was 
proposed in Ref.\,\cite{PAPIII} (see also Ref.\,\cite{rev}). Similarly as
the $B_s\to K\overline{K}$ strategy\,\cite{fd1} discussed in Subsection 3.3,
it makes use of the fact that the general phase structure of the 
corresponding decay amplitudes is known reliably within the Standard Model.  
If the magnitude of the current--current amplitude $T'$ 
contributing to $B^0_d\to\pi^-K^+$ is known -- it can be fixed similarly to 
$|T'_s|$, e.g.\ through $B^+\to\pi^+\pi^0$ and $SU(3)$ flavour symmetry, 
``factorization'', or hopefully lattice gauge theory 
one day -- two amplitude triangles can be constructed, allowing in particular
the extraction of $\gamma$. This approach is promising for future 
$B$-physics experiments, since it requires only time-independent measurements 
of branching ratios at the ${\cal O}(10^{-5})$ level. If one measures in 
addition the branching ratios for $B^+\to\pi^0K^+$ and its charge-conjugate, 
also the $b\to s$ electroweak penguin amplitude can be determined, which is -- 
among other things -- an interesting probe for new physics\,\cite{PAPI}.
 
Recently the CLEO collaboration\,\cite{cleo} has reported the first observation
of the decays $B^+\to\pi^+K^0$ and $B^0_d\to\pi^-K^+$. At present 
only combined branching ratios, i.e.\ averaged over decays and their 
charge-conjugates, are available with large experimental uncertainties. 
Therefore it is not yet possible to extract $\gamma$ from the triangle 
construction proposed in Ref.\,\cite{PAPIII}. The recent CLEO measurements 
allow us, however, to derive {\it constraints} on $\gamma$, which are of the 
form 
\begin{equation}\label{gamma-bound}
0^\circ\leq\gamma\leq\gamma_0\quad\lor\quad180^\circ-
\gamma_0\leq\gamma\leq180^\circ 
\end{equation}
and are hence complementary to the presently allowed range 
\begin{equation}\label{UT-fits}
42^\circ\mathrel{\hbox{\rlap{\hbox{\lower4pt\hbox{$\sim$}}}\hbox{$<$}}}
\gamma\mathrel{\hbox{\rlap{\hbox{\lower4pt\hbox{$\sim$}}}\hbox{$<$}}}135^\circ
\end{equation}
for that angle arising from the usual fits of the unitarity 
triangle\,\cite{bf-rev}. This interesting feature has recently been pointed 
out in Ref.\,\cite{fm2}. The quantity $\gamma_0$ in Eq.\ 
(\ref{gamma-bound}) depends both on 
\begin{equation}\label{Def-R}
R\equiv\frac{\mbox{BR}(B_d\to\pi^\mp K^\pm)}{\mbox{BR}(B^\pm\to\pi^\pm K)} 
=\frac{\mbox{BR}(B_d^0\to\pi^- K^+)+\mbox{BR}(\overline{B_d^0}\to\pi^+ 
K^-)}{\mbox{BR}(B^+\to\pi^+K^0)+\mbox{BR}(B^-\to\pi^-\overline{K^0})}\,,
\end{equation}
i.e.\ the ratio 
of combined branching ratios, and on the amplitude ratio $r\equiv|T'|/|P'|$
of the current--current and penguin operator contributions to $B_d\to\pi^\mp 
K^\pm$. 
 
While the constraints on $\gamma$ require knowledge about $r$ for $R>1$, 
bounds on $\gamma$ can always be obtained {\it independently} of $r$,
if $R$ is found experimentally to be smaller than 1. The point is that 
$\gamma_0$ takes a maximal value 
\begin{equation}
\gamma_0^{\rm max}=\mbox{arccos}(\sqrt{1-R})\,,
\end{equation}
which depends only on the ratio $R$ of combined $B_{u,d}\to\pi K$ branching 
ratios\,\cite{fm2}. 
 
Let us take as an example the central value 0.65 of the recent CLEO
measurements\,\cite{cleo}, yielding $R=0.65\pm0.40$. This value corresponds to 
$\gamma_0^{\rm max}=54^\circ$ and implies the range $0^\circ\leq\gamma\leq
54^\circ$ $\lor$ $126^\circ\leq\gamma\leq180^\circ$, which has only the small
overlap 
$42^\circ\mathrel{\hbox{\rlap{\hbox{\lower4pt\hbox{$\sim$}}}\hbox{$<$}}}
\gamma\leq54^\circ$ $\lor$ $126^\circ\leq\gamma
\mathrel{\hbox{\rlap{\hbox{\lower4pt\hbox{$\sim$}}}\hbox{$<$}}}135^\circ$ 
with the range (\ref{UT-fits}). The two pieces of this
range are distinguished by the sign of $\cos\delta$, where 
$\delta$ is the CP-conserving strong phase shift between the $T'$ and $P'$ 
amplitudes. Using arguments based on ``factorization'', one expects 
$\cos\delta>0$ corresponding to the former interval of that range\,\cite{fm2} 
(for a recent model calculation, see Ref.\,\cite{ag}). Consequently, once 
more data come in confirming $R<1$, the decays $B_d\to\pi^\mp K^\pm$ 
and $B^\pm\to\pi^\pm K$ may put the Standard Model to an interesting test and 
could open a window to new physics. Effects of physics beyond the Standard 
Model in $B_{u,d}\to\pi K$ decay amplitudes have been analysed in a recent 
paper\,\cite{fm3}. A detailed study of various implications of the bounds 
on $\gamma$ discussed above has recently been performed in Ref.\,\cite{gnps}, 
where the issue of new physics has also been addressed.

\subsection{Towards Extractions of $\gamma$ from $B_{u,d}\to\pi K$ and 
Untagged $B_s\to K\overline{K}$ Decays}\label{bsgam}
At first sight, the observables $a$ and $b$ specified in Eq.~(\ref{Def-ab}), 
which can be extracted from the time evolutions of the untagged 
$B_s\to K^0\overline{K^0}$ and $B_s\to K^+K^-$ decay rates, provide 
a similar bound on $\gamma$ as the combined $B_{u,d}\to\pi K$ 
rates. The point is that $R_s\equiv a+b$ corresponds exactly to the ratio $R$ 
of combined $B_{u,d}\to\pi K$ branching ratios.
 
A closer look shows, however, that it is possible to derive a more elaborate 
bound on $\gamma$ from the untagged $B_s\to K\overline{K}$ decay 
rates\,\cite{BpiKBsKK}:
\begin{equation}\label{cot-range}
\frac{\left|\,1-\sqrt{a}\,\right|}{\sqrt{b}}\leq|\cot\gamma\,|\leq\frac{1+
\sqrt{a}}{\sqrt{b}}\,,
\end{equation}
which corresponds to the allowed range 
\begin{equation}\label{Bs-bounds}
\gamma_1\leq\gamma\leq\gamma_2\quad\lor\quad180^\circ-\gamma_2\leq\gamma
\leq180^\circ-\gamma_1
\end{equation}
with
\begin{equation}
\gamma_1=\mbox{arccot}\left(\frac{1+\sqrt{a}}{\sqrt{b}}\right)\,,\quad
\gamma_2=\mbox{arccot}\left(\frac{\left|\,1-\sqrt{a}\,\right|}{\sqrt{b}}
\right)\,.
\end{equation}
It can be shown\,\cite{BpiKBsKK} that the bound Eq.~(\ref{Bs-bounds}) 
always excludes a range around $\gamma=90^\circ$ larger than the bound 
$|\cos\gamma\,|\geq\sqrt{1-R_s}$ arising for $R_s<1$, provided 
$R_s\not=\sqrt{a}$. For $R_s=\sqrt{a}$, both bounds exclude the same 
region around $\gamma=90^\circ$.

So far, we have discussed the $B_{u,d}\to\pi K$ and $B_s\to K\overline{K}$
decays separately. As we have just seen, interesting and stringent bounds
on $\gamma$ may arise from the corresponding observables. The goal is, 
however, not only to constrain $\gamma$, but to {\it determine} this angle 
eventually. If we consider $B_{u,d}\to\pi K$ and $B_s\to K\overline{K}$ 
separately, information about colour-allowed $\bar b\to\bar uu\bar s$ 
current--current amplitudes is needed to accomplish this 
task\,\cite{fd1,PAPIII}, which introduces hadronic uncertainties into the
extracted values of $\gamma$. Such an input can be avoided by relating
$r_s$ and $r$ straightforwardly through
\begin{equation}\label{SU3-break}
\left(\frac{r_s}{r}\right)^2\equiv\zeta\,,
\end{equation}
where $\zeta$ parametrizes $SU(3)$ breaking related to the $s$ and $u$, $d$
spectator quarks in the modes $B_{s}\to K\overline{K}$ and $B_{u,d}\to\pi K$,
respectively. As has recently been pointed out in Ref.\,\cite{BpiKBsKK},
the observables of the $B_{u,d}\to\pi K$ and $B_s\to K\overline{K}$ modes
allow a simultaneous determination of $\gamma$ and $r$, $\delta$, $r_s$,
$\delta_s$ (up to certain discrete ambiguities) as a function of the
$SU(3)$ breaking parameter $\zeta$. Keeping $\zeta$ explicitly in the 
formulae presented in Ref.\,\cite{BpiKBsKK} will turn out to be useful 
once $SU(3)$ breaking can be controlled in a quantitative way. As a 
first ``guess'', we may use $\zeta=1$, which appears to be a rather mild 
$SU(3)$ assumption, since $\zeta$ describes $SU(3)$ breaking originating 
only from different spectator quarks, which should play a minor role for 
the decay dynamics. One may also vary $\zeta$ within a reasonable range 
to get some feeling for the uncertainties arising from possible 
$SU(3)$ breaking. 

It is interesting to note that the approach sketched above works also for 
vanishing strong phases, and for a particular scenario of new physics, 
where the $B_{s}\to K\overline{K}$ and $B_{u,d}\to\pi K$ decay amplitudes are 
dominated by the Standard Model diagrams and $B^0_s$--$\overline{B^0_s}$ 
mixing receives CP-violating new-physics contributions. The observables 
of untagged $B_s\to K^0\overline{K^0}$ and $B_s\to D_s^+D_s^-$ decays 
provide an interesting test of whether such a scenario of physics beyond 
the Standard Model is actually realized in nature\,\cite{BpiKBsKK}. 

\section{Concluding Remarks and Outlook}
In conclusion, we have seen that the kaon system -- the only one where CP 
violation has been observed to date -- cannot provide the whole picture of 
that phenomenon. In addition to other interesting systems, for instance the 
$D$ system, non-leptonic $B$-meson decays, where large CP asymmetries are 
expected within the Standard Model, are extremely promising to test 
the CKM picture of CP violation. More advanced experimental studies of
CP-violating effects in the kaon system and the exploration of CP violation
at $B$ physics facilities are just ahead of us. In the foreseeable future,
these experiments may bring unexpected results that could shed light on the
physics beyond the Standard Model. Certainly an exciting future lies ahead
of us!

\vspace*{0.6cm}
\par\noindent
{\bf Acknowledgements}
\vspace*{0.3cm}
\par
I am very grateful to Michael Beyer and Thomas Mannel for inviting me to
such an interesting and well-organized workshop, and for providing generous
travel support.

\vspace*{0.6cm}
\par

\end{document}